\documentclass[a4paper,11pt]{article}
\pdfoutput=1

\usepackage{jcappub}

\usepackage[T1]{fontenc}

\usepackage{pstricks}
\usepackage{color}
\usepackage{graphicx}
\usepackage{multirow}
\usepackage[normalem]{ulem}

\usepackage{booktabs}

\newcommand{\eq}[1]{eq.~(\ref{eq:#1})}
\newcommand{\Eq}[1]{Eq.~(\ref{eq:#1})}
\newcommand{\secs}[1]{Section~\ref{sec:#1}}

\title{\boldmath Decaying Leptophilic Dark Matter at IceCube}

\author[a]{Sofiane M. Boucenna,}
\author[b,c]{Marco Chianese,}
\author[c]{Gianpiero Mangano,}
\author[b,c]{Gennaro Miele,}
\author[b,c]{Stefano Morisi,}
\author[b,c]{Ofelia Pisanti,}
\author[b]{and Edoardo Vitagliano}

\affiliation[a]{INFN, Laboratori Nazionali di Frascati, C.P. 13, I-00044 Frascati, Italy.}
\affiliation[b]{Dipartimento di Fisica, Universit\'a di Napoli ``Federico II'', Complesso Univ. Monte S. Angelo, Via Cinthia, I-80126 Napoli, Italy}
\affiliation[c]{INFN, Sezione di Napoli, Complesso Univ. Monte S. Angelo, Via Cinthia, I-80126 Napoli, Italy}
\emailAdd{boucenna@lnf.infn.it}
\emailAdd{chianese@na.infn.it}
\emailAdd{mangano@na.infn.it}
\emailAdd{miele@na.infn.it}
\emailAdd{stefano.morisi@gmail.com}
\emailAdd{pisanti@na.infn.it}
\emailAdd{ed.vitagliano@studenti.unina.it}

\abstract{
We present a novel interpretation of IceCube high energy neutrino events (with energy larger than 60 TeV) in terms of an extraterrestrial flux due to  two different contributions: a flux originated by known astrophysical sources and dominating IceCube observations up to few hundreds TeV, and a new flux component where the most energetic neutrinos come from the leptophilic three-body decays of dark matter particles with a mass of few PeV. Differently from other approaches, we provide two examples of elementary particle models that do not require extremely tiny coupling constants. We find the compatibility of the theoretical predictions with the IceCube results when the astrophysical flux has a cutoff of the order of 100 TeV (broken power law). In this case the most energetic part of the spectrum (PeV neutrinos) is due to an extra component such as the decay of a very massive dark matter component. Due to the low statistics at our disposal we have considered for simplicity the equivalence between deposited and neutrino energy, however such approximation does not affect dramatically the qualitative results. Of course, a purely astrophysical origin of the neutrino flux (no cutoff in energy below the PeV scale - unbroken power law) is still allowed. If future data will confirm the presence of a sharp cutoff above few PeV this would be in favor of a dark matter interpretation.}

\begin{document} 
\maketitle
\flushbottom

\section{Introduction}

After more than 80 years from its first evidence in the Coma galaxy cluster by Fritz Zwicky, the nature of Dark Matter (DM) still remains an open question. Elementary particle physics can provide interesting schemes where to allocate viable DM candidates, one of the most attractive and simple scenarios being the Weakly Interacting Massive Particle (WIMP) scenario with a DM mass in the range $\mathcal{O}(1)$~GeV~--~$\mathcal{O}(100)$~TeV \cite{Griest:1989wd}\footnote{This upper limit arises from a model independent unitary constraint. However, a typical mass for WIMP is $\sim \mathcal{O}(1)$~TeV \cite{Cirelli:2005uq}.} and interaction rates of the order of weak interactions. Even though such schemes naturally emerge in the SUSY extension of the electroweak Standard Model (SM),  up to now almost all  indirect or direct searches have not provided any clear evidence \cite{Agashe:2014kda}, and DM observations remain linked to their indirect gravitational footprint only.

Indeed, a large amount of different theoretical frameworks has been proposed in literature, where the mass of DM candidates is spread over many order of magnitude, from about $10^{-32}$  GeV up to $10^{18}$ GeV, e.g. axions, KeV sterile neutrinos,  majorons, or the heaviest wimpzilla ($\sim 10^{12}$ GeV). Lacking any direct DM detection at LHC experiments, it is likely that the only viable way to look for very massive DM candidates would exploit indirect searches in astrophysical observations. From this point of view, neutrino telescopes like IceCube (IC) provide a chance to observe high energy cosmic ray phenomena induced by such massive DM particles, where energetic neutrinos are produced.

The IceCube Neutrino Observatory experiment \cite{Aartsen:2013jdh,Aartsen:2014gkd} is an excellent example of high energy neutrino astronomy. This rapidly developing branch of physics is important for different reasons \cite{Gaisser:1990vg,Halzen:2002pg,Stanev:2004ys}. In fact, by looking at neutrinos, which are neutral and weakly interacting particles, one can  trace back the sources even in presence of an intergalactic background and magnetic fields. Moreover, the observation of astrophysical neutrinos  is of paramount importance, both because their presence is a proof that acceleration of hadronic matter is involved, and because the features in their energy spectrum may indicate their production in a top-down framework from interactions involving heavy non standard particles.

During three years (2010-2013) IC Collaboration has observed several neutrino events in the TeV-PeV range.  While the lower energy ones can be well explained in terms of atmospheric neutrinos, events in the range between 100 TeV and 2 PeV seem to be due to some extraterrestrial process. There are many astrophysical possible sources for such neutrinos \cite{Murase:2014tsa}, both from $pp$ \cite{Loeb:2006tw,Murase:2013rfa,Kelner:2006tc} and $p\gamma$ \cite{Mucke:1999yb} interactions; the bottom-up scenarios range from extragalactic Supernova Remnants (SNR) \cite{Chakraborty:2015sta} to Active Galactic Nuclei (AGN) \cite{Kalashev:2014vya,Stecker:1991vm} and Gamma Ray Bursts (GRB) \cite{Waxman:1997ti}, all of these with specific emission spectra due to the different production environment. For example, $p\gamma$ spectrum is peaked \cite{Mucke:1999yb} while $pp$ spectrum is flatter \cite{Kelner:2006tc}. According to the IC analysis \cite{Aartsen:2014gkd}, galactic sources alone can't explain the excess and extragalactic sources are necessary. However, even assuming an extragalactic origin, it is not straightforward to fit all the data; for example, $p\gamma$ AGN spectrum gives a good description of high energy  events but does not satisfactory fit lower energy data \cite{Stecker:1991vm}. On the other side, GRB predicted spectra agree with data modulo the normalization of flux, which, unfortunately, has an upper limit (given by searches for correlation with observed GRB) more than one order of magnitude below the observed value \cite{Abbasi:2012zw}.

More recently it has been proposed that high energy IC events could be related to the decay of DM particles \cite{Feldstein:2013kka,Esmaili:2013gha,Bai:2013nga,Ema:2013nda,Esmaili:2014rma,Bhattacharya:2014vwa, Higaki:2014dwa,Rott:2014kfa,Ema:2014ufa, Murase:2015gea,Dudas:2014bca,Fong:2014bsa}. Note that the presence of PeV decaying DM component can also alleviate some tension among cosmological parameters estimates \cite{Anchordoqui:2015lqa}. Such an interpretation would be supported by a lack of events above 2 PeV compatible with the decay of a massive particle with a mass of this order of magnitude, and by a distribution of arrival directions apparently uncorrelated with the galactic disk. If this interpretation is correct, IceCube will provide information about the DM mass range and cross-section with ordinary particles and thus, important hints for future DM experiments. 

For an elementary particle with a mass of $\mathcal{O}$(1)~PeV, the maximal annihilation cross-section obtained by saturating the unitarity bound yields to a completely negligible signal at IceCube. One is then compelled to consider decaying DM instead \cite{Feldstein:2013kka}.

Alternative connections between DM and IC events have been proposed in literature. In particular the boosted DM mechanism \cite{Bhattacharya:2014yha,Agashe:2014yua,Berger:2014sqa,Kopp:2015bfa} is based on the idea that a highly energetic (boosted) population of DM particle is originated by the decay of more massive and long-lived non-thermal relic, dominating the DM distribution. Such boosted particles then interact with nucleons of detector via neutral current interactions. Finally, different schemes have been proposed, where the bump in the neutrino flux at PeV observed by IceCube is explained as the s-channel enhancement of neutrino-quark scattering by a leptoquark with a mass of 0.6 TeV that couples to the $\tau$-flavour and light quarks \cite{Barger:2013pla}. 

In the present paper, we propose an interpretation of IceCube PeV events in terms of leptophilic three-body decays of a DM particle. The operator responsible for the decay results to have quite a {\it natural} coupling and is selected out by global flavour symmetries. Such symmetries forbid at the same time the two body decays and other dangerous three body-decay operators.

The paper is organized as follow. In \secs{secICdata} we give a brief review of IC data, and in \secs{sec:th} we outline the theoretical framework for DM particle. In \secs{DMflux} we discuss the neutrino flux produced, and in \secs{analysis}  we describe our results. \secs{concl} contains our conclusions.

\section{IC data}\label{sec:secICdata}

The IceCube Neutrino Observatory has been searching during three years (2010-2013) \cite{Aartsen:2014gkd} for astrophysical neutrinos in the energy range from 30 TeV to 100 PeV. The observed data have been analyzed in terms of their energy spectrum, arrival direction, and flavour. The expected background arises from muons and neutrinos coming from the decays of $\pi$ and $K$ produced by cosmic ray interactions in the atmosphere. Actually, when the energy of the parent meson increases, its lifetime becomes longer and correspondingly, the interaction probability dominates on the decay, giving a suppression in the atmospheric muon and neutrino flux at high energy. No suppression is instead foreseen for the atmospheric neutrino background coming from the decay of charmed mesons, the so-called {\it prompt} component, because of the short lifetime of these mesons. Measuring the muon detection rate in a separate region of the telescope, the IC collaboration gives the following estimation of the muon and neutrino background
\begin{equation}
N_{\mu^\pm}=8.4\pm4.2 ~~~~~~~~~~~~~~~ N^{all}_{\nu+\bar{\nu}}=6.6^{+5.9}_{-1.6} \ ,
\end{equation}
where $N^{all}_{\nu+\bar{\nu}}$ stands for the number of all flavour neutrinos and antineutrinos and its asymmetric error is due to the prompt component.

IceCube collected 37 events in 988 days, with deposited energies ranging from 30 TeV to 2 PeV. In particular, three events were detected, with deposited energy of the order of PeV, which are the most energetic neutrino events ever detected. Among all the events, two of them, events 28 and 32, having sub-threshold signals in IceTop, seem to be part of the expected muon background (in particular, event 32 cannot be reconstructed with a single direction and energy), while three ambiguous downgoing tracks seem to be of an atmospheric origin. 

It has been argued in literature \cite{Lipari:2013taa}  that, due to the present uncertainty on the overall rate and starting energy of the prompt neutrino component, by assuming a different cross section for charmed meson production and/or a slightly different cosmic ray primary flux one could modify the expected atmospheric neutrino background in such a way to reduce the necessity of an extra neutrino flux in the high energy range. In this case since about 50\% of downgoing prompt neutrinos arrive together with muons, which should trigger the muon veto, upgoing events should be more than downgoing ones at these energies (``southern hemisphere'' suppression). IceCube observes exactly the contrary, so it presently seems unlikely that the atmospheric background and the observed neutrino flux could be reconciled thanks to prompt neutrinos \cite{Gaisser:2014bja,Bhattacharya:2015jpa}. 

Another indication comes from the topology of detected events, which belong to two classes: track events, associated with the propagation of a high energy muon, and shower events, which correspond to the production of a large aggregate of secondary particles (with similar energy resolution in the two cases but, of course, better angular resolution in the first one). By assuming that data are due to a purely conventional atmospheric flux, one should count more tracks than showers (since atmospheric neutrinos are mainly muon neutrinos). On the other side, a $(1/3:1/3:1/3)$ flavour proportion in the neutrino flux, as expected for astrophysical neutrinos, would result in only 20\% $\nu_\mu$ CC interactions, a closer result to the IceCube finding of less tracks (24\%) than showers (76\%).

On the basis of this analysis the IC collaboration concludes \cite{Aartsen:2014gkd} that a purely atmospheric origin of the high energy events, requiring a 3.6 times higher charm normalization, is rejected at 5.7$\sigma$. Moreover, the data are well described in terms of a global fit including background atmospheric muons and neutrinos, prompt neutrinos and an isotropic astrophysical flux, which for each flavour takes the form (quoted errors are 1-$\sigma$ uncertainties)
\begin{equation}
E^2\frac{dJ_{\nu+\bar{\nu}}}{dE}=\left(0.95\pm0.3\right)\times 10^{-8}\mbox{GeV}\mbox{ cm}^{-2}\mbox{ s}^{-1}\mbox{ sr}^{-1} \ .
\end{equation}
This result satisfies the Waxman-Bahcall bound for optically thin sources \cite{Waxman:1998yy}, obtained supposing that all the charged particles created by cosmic accelerators give their energy to kaons and pions.

While the unbroken $E^{-2}$ hypothesis gives a fairly good description of the data energy spectrum, it predicts 3.1 more events at 2 PeV, seeming to require a softer spectrum or a high energy cutoff. A more general model of the astrophysical component by a piecewise function of the energy gives the best-fit
\begin{equation}
E^2\frac{dJ_{\nu+\bar{\nu}}}{dE}=1.5\times 10^{-8}\left(\frac{E}{100 \mbox{ TeV}}\right)^{-0.3}\mbox{GeV}\mbox{ cm}^{-2}\mbox{ s}^{-1}\mbox{ sr}^{-1} \ .
\end{equation}  
This corresponds to the lower boundary of the total statistical and systematic uncertainty on the energy power law (with a zero charm contribution), which on the other side reaches, at 90\% C.L., the previous $E^{-2}$ behaviour.

In the present work, by following the background analysis of ref. \cite{Aartsen:2014gkd}, we assume that neutrino flux is mainly dominated by the {\it atmospheric component} up to 60 TeV, whereas we consider a bottom-up neutrino contribution from known astrophysical sources (like for example extragalactic SNR) in the [60 TeV, 300 TeV] range (hereafter denoted as {\it astrophysical neutrino flux}), and, at the same time, a top-down additional component at higher energy (hereafter denoted {\it DM neutrino flux}). This would naturally produce a sharp cutoff, observed in the data around few PeV, and if confirmed by new data, represents an intriguing feature of the IC observations.

\section{The Theoretical Framework}
\label{sec:sec:th}

For a heavy fermion singlet DM candidate, $\chi$, which is directly coupled to neutrinos, the lowest dimension coupling is a Yukawa interaction \begin{equation}\label{eqLnC}
y\, \bar{L}_\alpha \tilde{\phi} \chi \ , 
\end{equation}
where $\tilde\phi\equiv i\sigma_2 \phi^*$ ($\phi$ denoting the SM Higgs doublet), $L_{\alpha=e,\mu,\tau}$ stand for the lepton doublets, and $\sigma_2$ is a Pauli matrix. As shown in \cite{Esmaili:2014rma}, this new interaction term is cosmologically safe (sufficiently long lived DM), and relevant for IceCube (a DM particle specie abundant enough to fit the observed flux) for a {\it fine tuned} tiny coupling, $y \sim \mathcal{O}(10^{-30})$. Note that the interaction term in Eq.~\eqref{eqLnC} is reminiscent of the right-handed neutrino in see-saw models, but its  contribution to the lightest neutrino mass would be negligible due to the smallness of the $y$ coupling. The operator in Eq.\,(\ref{eqLnC}) would yield a sharp peak in energy. The presence of the Higgs field in Eq. (\ref{eqLnC}) allows for an abundant production of secondary neutrinos via the decay of the Higgs particles to heavy quarks, giving an almost flat neutrino flux at energies lower than PeV \cite{Esmaili:2014rma}.  Such a result could turn out to be problematic if known astrophysical contributions were included in the analysis. For instance in Ref.~\cite{Chakraborty:2015sta}, low energy (up to 100 TeV) IC events are shown to be well fitted by means of standard extragalactic SNR. 

In the present paper we try both to improve the need of an unnatural coupling and at the same time, to reproduce the IC data including an astrophysical neutrino component. This suggests to consider an interaction term that does not directly involve quarks, the Higgs field or gauge bosons (leptophilic DM), and which is a higher dimension operator. In this case, the feeble coupling can be understood in terms of a large mass scale. In particular we assume that:
\begin{itemize}
\item[1)] the DM field $\chi$ is coupled to the SM particles via a leptophilic coupling;
\item[2)] the lifetime of $\chi$ is suppressed by powers of the scale of new physics entering in the non-renormalizable coupling;
\item[3)] there is a direct coupling to neutrinos allowing for a primary neutrino flux with energy of the order of $\chi$ mass. The multi-body final state may contain lower energy neutrinos as well, so that neutrino flux also spreads to lower energies;
\item[4)] mass and couplings of $\chi$ are determined in order to produce a neutrino flux just dominating the energy region around PeV. This implies, on the contrary, that the flux in the energy region [60 TeV, 300 TeV] is due to astrophysical sources as will be discussed in the following. We also assume for simplicity that $\chi$ represents the dominating contribution of  Cold DM.
\end{itemize}
As in~\cite{Haba:2010ag}, one can list the gauge-invariant operators up to dimension--6, shown in Table \ref {op}. $L$ and $Q$ are the lepton and quark weak doublets and  $u$ and $d$ the corresponding right-handed quarks. The field ${\ell}$ stands for the right-handed lepton singlet, and finally, $W_{\mu \nu}$ and $B_{\mu \nu}$ are the field strength tensors of $SU(2)_L$ and $U(1)_Y$ gauge bosons. To simplify notation, we have omitted the family index for matter fields.
\begin{table}[t!]
\centering
\begin{tabular}{|c|cl|} \hline
Dimensions && \multicolumn{1}{c|}{DM decay operators} \\ \hline \hline
&&\\
4 && $\bar{L} \tilde{\phi} \chi$ \\ 
5 && ~~~$-$ \\
6 && 
$\bar{L}{\ell} \, \bar{L}\chi$,
~~$\phi^\dagger\! \phi \bar{L}\tilde{\phi} \chi$,
~~$(\tilde{\phi})^tD_\mu \tilde{\phi} \bar{{\ell}}\gamma^\mu \chi$, \\[3mm]
&&
$\bar{Q}d\, \bar{L}\chi$,
~~$\bar{u}Q\, \bar{L}\chi$,
~~$\bar{L} d \, \bar{Q}\chi$,
~~$\bar{u}\gamma_\mu d\, \bar{{\ell}}\gamma^\mu \chi$, \\[3mm]
&& 
$D^\mu \tilde{\phi} D_\mu \bar{L} \chi$,
~~$D^\mu D_\mu \tilde{\phi} \bar{L}\chi$,  \\[3mm]
&& 
$B_{\mu\nu}\bar{L}\sigma^{\mu\nu}\tilde{\phi} \chi$,
~~$W_{\mu\nu}^a\bar{L}\sigma^{\mu\nu}\tau^a \tilde{\phi} \chi$  \\[3mm]
 \hline
\end{tabular}
\caption{Gauge-invariant operators up to dimension--6 inducing fermion singlet DM decay as taken from Table 1 of Ref.~\cite{Haba:2010ag}. The adopted notation is explained in the text.}
\label{op}
\end{table}

Remarkably,  the only operator in this list which satisfies requirements 1)-3) is the {\it non-renormalizable lepton portal}:
\begin{equation}
\label{eq:DMop}
\frac{y_{\alpha\beta\gamma}}{M_{\rm Pl}^2} \left(\overline{L_\alpha}  {\ell}_\beta\right) \left( \overline{L_\gamma} \chi \right)+ \mathrm{h.c.} \ ,
\end{equation}
where $\{\alpha,\beta,\gamma\}$ are flavour indices thus labeling 27 operators, and the round brackets indicate the Lorentz contractions. The mass scale here is chosen as the Planck mass $M_{\rm Pl}$. Indeed, for this choice the order of magnitude of couplings $y$ will not be unnaturally small, see \secs{analysis}. 

If this operator is the only source of DM decay, one has to invoke some selection rule which forbids the dimension--4 operator, which, as we mentioned, is compatible with IC results for an extremely small coupling only. This can be done by using global flavour symmetries, both Abelian like $U_f(1)$ and non-Abelian groups like $A_4$, $\Delta(27)$, etc. We will focus in the following on two benchmark schemes (see Appendix for more details about these models)\footnote{In our analysis we assume for simplicity that $\chi$ is a Dirac fermion. If $\chi$ was a Majorana fermion, one could not reproduce, for example, the Abelian  model 1) that requires a flavour charge for $\chi$ different from 0, whereas a scheme like $A_4$ would still be possible.}:
\begin{itemize}
\item model 1) - $U_f(1)$ symmetry \qquad $\{\alpha,\beta,\gamma\}\equiv \{\mu, e,\tau \}+\{\tau,e, \mu \}+\{e, \mu, e \}$;  
\item model 2) - $A_4$ symmetry \qquad ~~~  $\{\alpha,\beta,\gamma\}\equiv \{e,\mu,\tau  \}$ +  cyclic permutations; 
\end{itemize}
where the brackets show the flavour assignments corresponding to non vanishing couplings $y_{\alpha \beta \gamma}$. Note that expanding the $SU(2)$ contractions, the operator of Eq.~\eqref{eq:DMop} always yields a coupling of DM with two charged leptons and one neutrino. Depending on the flavour index  the charged leptons can then possibly decay producing  secondary neutrinos. 

\section{DM neutrino flux}\label{sec:DMflux}

Following the method outlined in Ref.~\cite{Esmaili:2014rma} we consider both the contributions to top-down neutrino flux coming from the galactic  and extragalactic distributions of DM particles $\chi$ (the sum of $\chi$ and $\bar{\chi}$ contributions is implicitly assumed)
\begin{equation}
\frac{{\rm d} J_{\chi}}{{\rm d}E_\nu} \left(E_\nu\right) = \frac{1}{4 \pi} \int d\Omega \left( \frac{{\rm d} J_{\chi}^{\rm G}}{{\rm d}E_\nu}(E_\nu,l,b)+ \frac{{\rm d} J_{\chi}^{\rm EG}}{{\rm d}E_\nu}\left(E_\nu\right) \right)\ ,
\label{eq_DMflux}
\end{equation}
where the solid angle integration is on the longitude and latitude in the galactic coordinate system, $l$ and $b$. In the following we will always sum the flux of neutrinos and antineutrinos of different flavours, unless explicitly stated.

The galactic component due to the Milky Way halo can be written as
\begin{equation}
\frac{{\rm d} J_{\chi}^{\rm G}}{{\rm d}E_\nu}(E_\nu,l,b) = 
\frac{1}{4\pi\,M_{\chi}\,\tau_{\chi}} 
\sum_{\alpha=e,\mu,\tau} \frac{{\rm d}N^\alpha_{\nu + \bar{\nu}}}{{\rm d}E_\nu}\left(E_\nu\right)
\int_0^\infty {\rm d}s\; 
\rho_{\chi}(r(s,l,b))\ ,
\end{equation}
where $M_{\chi}$ and $\tau_{\chi}$ denote the mass and lifetime of DM particle. The quantity $\rho_{\chi}(r)$ denotes the density profile of DM particles in our Galaxy as a function of distance from the Galactic center, $r$, and ${\rm d}N^\alpha_{\nu + \bar{\nu}}/{\rm d}E_\nu$ stands for the energy spectrum of neutrinos and antineutrinos of flavour $\alpha$ produced in the decay. Note that the parameter $s$ is related to $r$ via the expression
\begin{equation}
r(s,l,b) = \sqrt{s^2+R^2_\odot-2 s R_\odot \cos b\cos l} \ ,
\end{equation}
where $R_\odot\simeq 8.5\,{\rm kpc}$ is the distance of the Sun from the Galactic center. In the following we assume a Navarro-Frenk-White density profile for the halo 
\begin{equation}
\rho_{\rm \chi}(r)= \frac{\rho_\chi}{r/r_c (1+r/r_c)^2}\ ,
\end{equation}
with $r_c =20\,\text{kpc}$ the critical radius and $\rho_\chi = 0.33\,{\rm GeV}\, {\rm cm}^{-3}$. We have checked the dependence of our predictions on the choice for the density profile (Einasto, Isothermal, etc.). The results are very similar, with a variation of the best fit values of the order of few percents. 

The quantity ${\rm d}N^\alpha_{\nu + \bar{\nu}}/{\rm d}E_\nu$ has been evaluated by means of a MonteCarlo procedure. In our code we take into account primary neutrinos (antineutrinos) and also the neutrinos (antineutrinos) produced by the decay of $\mu$ and $\tau$ leptons. The $\tau$ leptons have different decay channels that involve pions as well \cite{Agashe:2014kda}. The MonteCarlo takes into account $\sim 90 \%$ of the $\tau$ decay width, including the two leptonic decays into muons and electrons, and semileptonic decay channels up to three pions in the final state, whose charged states eventually decay mainly producing $\mu$ and corresponding neutrinos (antineutrinos). It is worth observing that the electroweak radiative corrections produce a large number of secondary neutrinos that are typically placed at energies almost two orders of magnitude smaller than the mass of DM particle \cite{Ciafaloni:2011sa}. Hence, for a mass of few PeV, such low energy neutrinos do not significantly affect our analysis that is restricted to neutrino energies larger than 60 TeV. In our study, we also consider the total energy injected by DM decay in the electromagnetic (e.m.) sector, and compare such a value with the bound provided by Fermi-LAT \cite{Ackermann:2014usa}.

The isotropic extragalactic component of the differential flux is given by 
\begin{equation}
\frac{{\rm d}J^{\rm EG}_\chi}{{\rm d}E_\nu} \left(E_\nu\right) =
\frac{\Omega_{\chi}\rho_{\rm cr}}{4\pi M_{\chi} \tau_{\chi}}
\int_0^\infty {\rm d}z\,
\frac{1}{H(z)}
\sum_{\alpha=e,\mu,\tau} \frac{{\rm d}N^\alpha_{\nu + \bar{\nu}}}{{\rm d}E_\nu}\left((1+z)E_\nu\right),
\end{equation}
where $H(z)=H_0 \sqrt{\Omega_\Lambda+\Omega_{\rm m}(1+z)^3}$ is the Hubble expansion rate as a function of redshift $z$ and $\rho_{\rm cr}=5.5\times10^{-6}\,{\rm GeV}\, {\rm cm}^{-3}$ is the critical density of the Universe. We assume a $\Lambda$CDM cosmology with parameters $\Omega_\Lambda=0.6825$, $\Omega_{\rm m}=0.3175$, $\Omega_{\chi}=0.2685$ and $h\equiv H_0/100\,{\rm km}\,{\rm s}^{-1}\,{\rm Mpc}^{-1}=0.6711$, according to Planck experiment results \cite{Ade:2015xua}. We found that the galactic and extragalactic components of the DM differential flux are of the same order of magnitude.

The $\chi$-lifetime, $\tau_\chi$, depends on the values assumed for $y_{\alpha\beta\gamma}$. For models 1) and 2) we get 
\begin{eqnarray}
{\rm model \, \, 1)} \qquad
\tau_\chi^{-1} & = &\frac{1}{6144 \, \pi^3} \left( 2 \left| y_{\mu e \tau} - y_{\tau e \mu} \right|^2 +  \left| y_{e \mu e} \right|^2 \right) \frac{M_\chi^5}{M_{\rm Pl}^4} \ ,\\
{\rm model \, \, 2)} \qquad
\tau_\chi^{-1} & = &\frac{1}{1024 \, \pi^3} \left( \left| y_+ \right|^2 + 3 \left| y_- \right|^2 \right) \frac{M_\chi^5}{M_{\rm Pl}^4} \ ,
\end{eqnarray}
where $y_+$ and $y_-$ denote the two independent symmetric and antisymmetric couplings  in $A_4$ \cite{Haba:2010ag}. In the previous expressions we neglected all final state masses,  due to the large value of $M_\chi$.

In the following, for simplicity we assume real couplings and the relations
\begin{eqnarray}
{\rm model \, \, 1)}  &\qquad & \left| y_{\mu e \tau} - y_{\tau e \mu} \right|=\left| y_{e \mu e} \right| \equiv y \label{eq_def_y_1} \ , \\
{\rm model \, \, 2)}  &\qquad &  \left|y_+\right|=\left|y_-\right| \equiv y \label{eq_def_y_2} \ .
\end{eqnarray}

\subsection{DM relic abundance}

As well known,  the standard thermal freeze-out mechanism is not a viable option for particles with masses larger than $\mathcal{O}(100)$~TeV because of the unitarity bound on the cross-section~\cite{Griest:1989wd}.  Hence, considering a DM candidate of PeV mass raises the question of the production mechanism, which is crucial to determine its relic abundance. There exist several scenarios, beyond the WIMP paradigm, which make such a heavy particle a perfectly viable DM candidate (see, e.g., Ref.~\cite{Baer:2014eja} for a review).

One possibility is to consider a non-thermal production during reheating of the Universe, after the inflationary stage, in a cosmological scenario with a low reheating temperature~\cite{Chung:1998rq,Giudice:2000ex}. After inflation, the Universe reaches its maximal temperature, $T_{max}$. This temperature can be much higher than the reheating temperature, $T_{RH}$, defined as
\begin{equation}
T_{RH} = 0.2 \left(\frac{200}{g_\star^{RH}}\right)^{1/4} \sqrt{\Gamma_\rho M_{\rm Pl}}\ ,
\end{equation}
where $g_\star^{RH} \equiv g_\star(T_{RH})$ is the effective number of relativistic degrees of freedom at $T_{RH}$, and $\Gamma_\rho$ the inflaton field decay rate. If the $\chi$ particle production takes place between $T_{max}$ and $T_{RH}$, the DM abundance scales as \cite{Chung:1998rq}:
\begin{equation}
\label{eq:DMab}
\Omega_\chi h^2 \sim M_\chi^{2} \langle\sigma v\rangle \left(\frac{g_\star^{RH}}{200}\right)^{-3/2}\,\left(\frac{2000\, T_{RH}}{M_\chi}\right)^7  \ ,
\end{equation}
where $\langle\sigma v\rangle$ is the thermal average of the DM  annihilation cross-section times the M{\o}ller flux factor. Notice that the $\chi$ particles are never in chemical equilibrium (even if they are in kinetic equilibrium), so their abundance is described by a power law instead of being exponentially suppressed as $\exp(-M_\chi/T_{RH})$. Assuming $\langle\sigma v\rangle \sim M_\chi^{-2}$, we obtain for $g_\star^{RH}= 200$ and $M_\chi = 1 \,\mathrm{PeV}$, 
\begin{equation}
 \label{eq:DMTR}
T_{RH}^{(1 \,\mathrm{PeV})} \approx 370 \,\mathrm{GeV} \ ,
\end{equation}
if $\chi$ provides the whole contribution to DM today. The previous equations are valid under the condition $T_{max} >M_\chi\gtrsim 20 \,T_{RH}$~\cite{Chung:1998ua}, corresponding to the fact that DM never reached local thermal equilibrium in the early Universe (which would otherwise spoil the result of \eq{DMab}). From \eq{DMTR} we see that the lower bound is readily satisfied.

Another viable possibility is the production through inelastic scattering interactions among the thermal plasma and energetic particles originating from inflaton decay and taking place at temperatures between $T_{max}$ and $T_{RH}$~\cite{Harigaya:2014waa}. In this case, assuming that DM couplings are of same order as gauge couplings and that $M_\rho>M_\chi^2/2 T_{RH}$, with $M_\rho$ the mass of the inflaton, in order to account for the observed DM abundance the  following condition should be satisfied~\cite{Harigaya:2014waa}:
\begin{equation}
\label{eq:DMplasma}
M_\chi \sim 1\, \mathrm{PeV}\, \left(\frac{T_{RH}}{80 \,\mathrm{MeV}}\right)^{3/2} \ .
\end{equation}
This mechanism would require quite low, thought viable, values for the reheating temperature to account for $M_\chi \sim 1$ PeV.

Finally, PeV dark matter could also be produced from inflaton decays (directly or through cascades). However, this mechanism is quite model-dependent because it strongly depends on the couplings of the inflaton to the particles of the model~\cite{Gelmini:2006pw,Allahverdi:2002nb}. For the sake of illustration, the direct production of DM from inflaton decay gives the abundance (assuming all decays happen at $T_{RH}$):
\begin{equation}
\Omega_\chi h^2 \approx 2.1 \times 10^8 \, \frac{M_\chi}{M_\rho}\frac{T_{RH}}{\mathrm{GeV}}\, B(\rho \to DM) \ ,
\end{equation}
where $B(\rho \to DM)$ is the average number of DM particles produced per inflaton decay.

\section{Analysis and results}
\label{sec:analysis}

In order to reproduce the sharp cutoff observed in the IC neutrino data, in the present analysis we assume that the neutrino flux is given by the combination of an atmospheric component up to 60 TeV \cite{Aartsen:2014gkd} and two different neutrino contributions, a bottom-up one from known astrophysical sources, like extragalactic SNR, in the [60 TeV, 300 TeV] range (hereafter denoted as $J_{\rm Ast}$), and a top-down component at higher energies (that is the flux $J_{\chi}$, defined in Eq.~\eqref{eq_DMflux}). Hence, for $ E_\nu \geq 60$ TeV the total neutrino flux is given by
\begin{equation}
\frac{{\rm d} J}{{\rm d}E_\nu} \left(E_\nu\right) =  \frac{{\rm d} J_{\chi}}{{\rm d}E_\nu} \left(E_\nu\right) + \frac{{\rm d} J_{\rm Ast}}{{\rm d}E_\nu} \left(E_\nu\right) \,.
\label{eq:flux_sum}
\end{equation}
Then, the number of neutrinos in a given energy bin $\left[ E_i,E_{i+1}\right]$ is equal to
\begin{equation}
N_i = 4 \pi \Delta t \int^{E_{i+1}}_{E_i} {\rm d}E \sum_{\alpha=e,\mu,\tau} \frac{{\rm d}J^\alpha_{\nu+\overline{\nu}}}{{\rm d}E} \,A_\alpha \left(E\right) \,,
\label{eq:number_neutrino}
\end{equation}
where $\Delta t=988$~days is the exposure time after 3-years of IC experiment, and $A_\alpha \left(E\right)$ is the neutrino effective area \cite{Aartsen:2013jdh} for different neutrino flavour $\alpha$. In order to compare the theoretical predictions of Eq. \eqref{eq:number_neutrino} with the observations we need to recast such expression in terms of deposited energy. In general, to statistically estimate the ratio between the deposited and neutrino energies a MonteCarlo simulation of the apparatus is required \cite{Aartsen:2013vja}. When for a bin of the deposited energy a significant  statistics is collected one could apply an average ratio that results to be of the order of $ (\sigma^{CC} \, 97 \% +  \sigma^{NC} \, 23 \%)/ (\sigma^{CC}+\sigma^{NC}) \sim 75 \%$ (see Table 1 of \cite{Aartsen:2013vja}). Remarkably, such a number appears to be quite stable as a function of the neutrino energy.  Unfortunately, due to low statistics collected till now, this procedure would be characterized by a large uncertainty in the energy determination. For this reason we prefer in this paper to assume the simplicity ansatz that the two energies coincide. Notice that in any case an expected shift in the energy of the order of 25$\%$ is not going to change dramatically our conclusions.

Known astrophysical sources able to produce a neutrino flux similar to the one observed by IceCube are high energy accelerators \cite{Murase:2014tsa} such as extragalactic SNR \cite{Chakraborty:2015sta}, Hypernova Remnants (HNR) \cite{Chakraborty:2015sta}, AGN \cite{Kalashev:2014vya,Stecker:1991vm}, and GRB \cite{Waxman:1997ti}. It is worth observing that, due to the large uncertainties in the parameters related to the physics of the accelerator mechanisms, there is enough room to reproduce IC neutrino flux. In particular, an extragalactic SNR can produce a neutrino flux with a cutoff ${\cal{O}}(100)$~TeV coming from the proton-proton hadronic interactions up to few PeV. On the other hand the HNR are sources of 100 PeV protons that can explain the PeV neutrino events \cite{Chakraborty:2015sta}. While classical GRB are ruled out, low-luminosity GRB and choked jets, which cannot be triggered by GRB satellites, are totally viable as the origin of PeV neutrinos \cite{He:2012tq,Murase:2013ffa,Tamborra:2015qza}. AGN can be the sources of PeV neutrinos \cite{Stecker:1991vm,Murase:2014foa}, but if the flux is normalized to match the observed IC PeV events, the lower energy part is inconsistent with data. All these astrophysical sources have also an associated $\gamma$-ray flux that would give a significant contribution to the diffuse $\gamma$-ray background, strongly constrained by Fermi-LAT \cite{Ackermann:2014usa}.

In order to parametrize the astrophysical flux one can use either an Unbroken Power Law (UPL), with a power law behavior in the whole IC region, or a Broken Power Law (BPL) where an exponential cutoff is assumed at some energy scale $E_0$. Considering both options essentially covers the wide range of accelerator mechanisms related to the different astrophysical sources. Hence, using the notation adopted by the IC Collaboration, we will consider:
\begin{itemize}
\item[i)] Unbroken Power Law (UPL): 
\begin{equation}
E_\nu^2 \frac{{\rm d} J_{\rm Ast}}{{\rm d}E_\nu} \left(E_\nu\right) = J_0 \,\left(\frac{E_\nu}{100\,  {\rm TeV}}\right)^{-\gamma} \,,
\label{eq_UPL_def}
\end{equation}
\item[ii)] Broken Power Law (BPL):  
\begin{equation}
E_\nu^2 \frac{{\rm d} J_{\rm Ast}}{{\rm d}E_\nu} \left(E_\nu\right)  = J_0 \,\left(\frac{E_\nu}{100\,  {\rm TeV}}\right)^{-\gamma}\, \exp{\left( -\frac{E_\nu}{E_0}\right)} \ ,
\label{eq_BPL_def}
\end{equation}
\end{itemize}
where $\gamma+2$ is the {\it spectral index} and  $J_0$ is the flux normalization. In the present analysis we fix the value of $E_0$ to be equal to $125$ TeV in agreement with the extragalactic SNR results \cite{Chakraborty:2015sta}. Furthermore, we restricted the spectral index to the physical range $\gamma \in [0, 1]$, as suggested by cosmic accelerator mechanisms.
\begin{table}[t!]
\begin{center}
\begin{tabular}{|c|c|c|c|c|}
\hline
Model& Case & $y \, [10^{-5}]$ & $J_0\, [10^{-8}]$ &  ${\chi}^2$/dof\\
\hline
\hline
\multicolumn{1}{ |l  }{\multirow{4}{*}{1) $U_f(1)$} } &
\multicolumn{1}{ |c| }{} & & &   \\
\multicolumn{1}{ |c  }{}                        &
\multicolumn{1}{ |c| }{UPL} &  1.0$^{+0.7}_{-0.7} $ & 0.8$^{+1.0}_{-0.5} $  &  $10.3/12$     \\[3mm]
\multicolumn{1}{ |c  }{}                        &
\multicolumn{1}{ |c| }{BPL} &  1.1$^{+0.6}_{-0.5} $ & 2.5$^{+2.8}_{-2.1} $ & $9.2/12$   \\[3mm]
\hline 
\multicolumn{1}{ |l  }{\multirow{4}{*}{2) $A_4$} } &
\multicolumn{1}{ |c| }{} & & &   \\
\multicolumn{1}{ |c  }{}                        &
\multicolumn{1}{ |c| }{UPL} & 0.35$^{+0.21}_{-0.21} $ &  0.8$^{+1.0}_{-0.5} $  & $10.7/12$    \\[3mm]
\multicolumn{1}{ |c  }{}                        &
\multicolumn{1}{ |c| }{BPL} & 0.37$^{+0.17}_{-0.16} $ &  2.4$^{+2.8}_{-2.0} $  & $9.6/12$      \\[3mm]
\hline
\end{tabular}
\caption{The marginalized 95\% C.L. for parameters $y$ and $J_0$ (expressed in unit of GeV cm$^{-2}$ s$^{-1}$ sr$^{-1}$) corresponding to models 1) and 2), and UPL and BPL parameterizations, respectively. The last column reports the reduced $\chi^2$.}
\label{tab2}
\end{center}
\end{table}

The fit has been done by means of a multi-Poisson likelihood analysis \cite{Baker:1983tu} in which the $\chi^2$ takes the expression
\begin{equation}
\chi^2=-2 \ln {\mathcal{L}}=2 \sum_i \left[N_i - n_i + n_i \ln\left(\frac{n_i}{N_i}\right) \right] \,,
\end{equation}
where $N_i$ is the expected number of neutrinos for energy bin provided by Eq.~\eqref{eq:number_neutrino}, while $n_i$ is the observed one. Once the atmospheric background has been subtracted, we fit the IC data by using the parametrization of the astrophysical flux in both cases of UPL and BPL, and considering model 1) and model 2) of \secs{sec:th} for the DM neutrino flux. The mass of DM particle has been varied in a range [1~PeV,10~PeV] able to produce a drop in the flux. The best fit value is found for $M_\chi = 5.0$ PeV independently of the model adopted. On the other side, after scanning the $\gamma$ range, we find a best fit value $\gamma=1.0$ for UPL (spectral index = 3.0), and $\gamma=0.0$ for BPL (spectral index = 2.0).
\begin{figure}[t!]
\begin{center}
\includegraphics[width=0.46\textwidth]{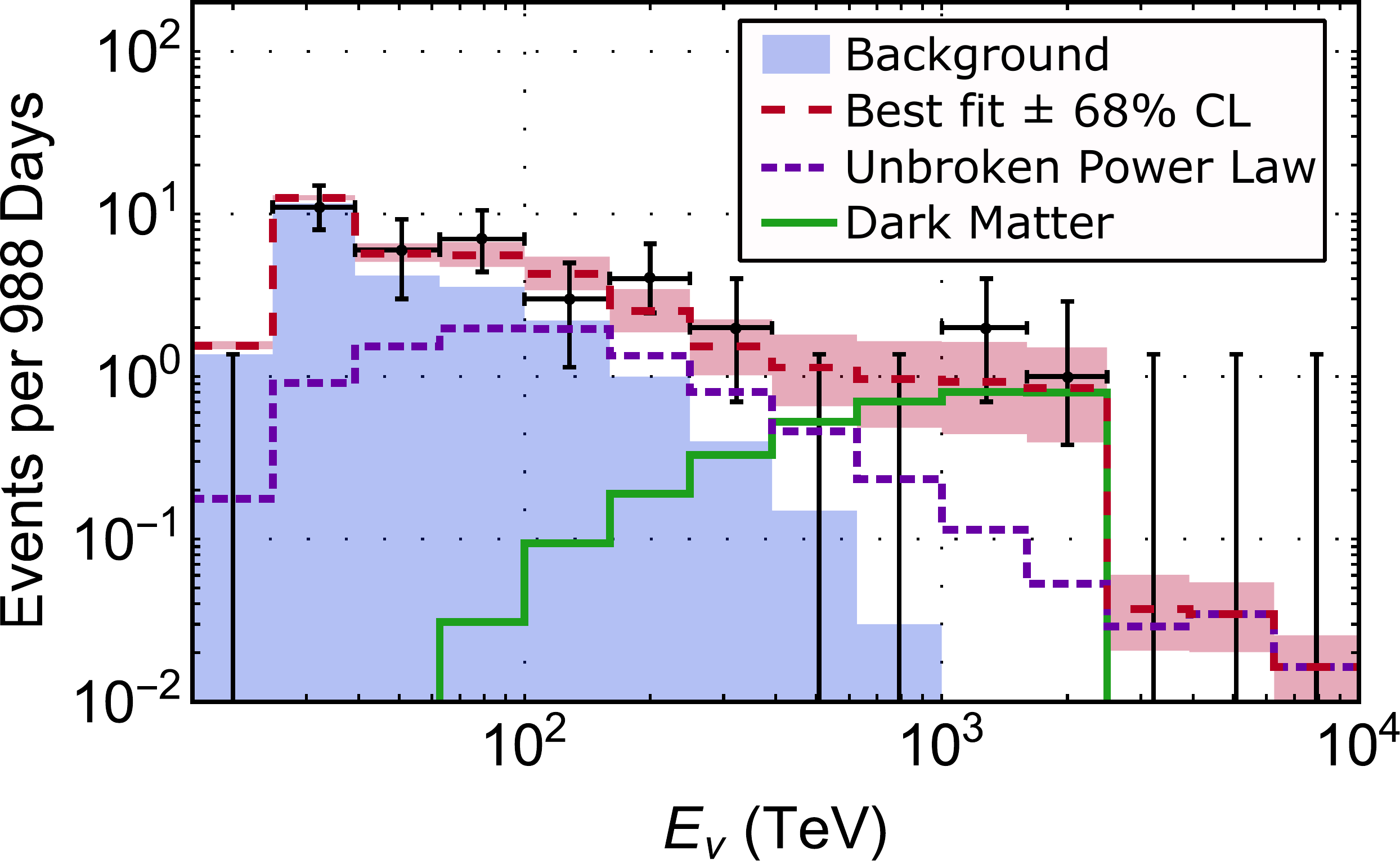}
\hskip5.mm
\includegraphics[width=0.46\textwidth]{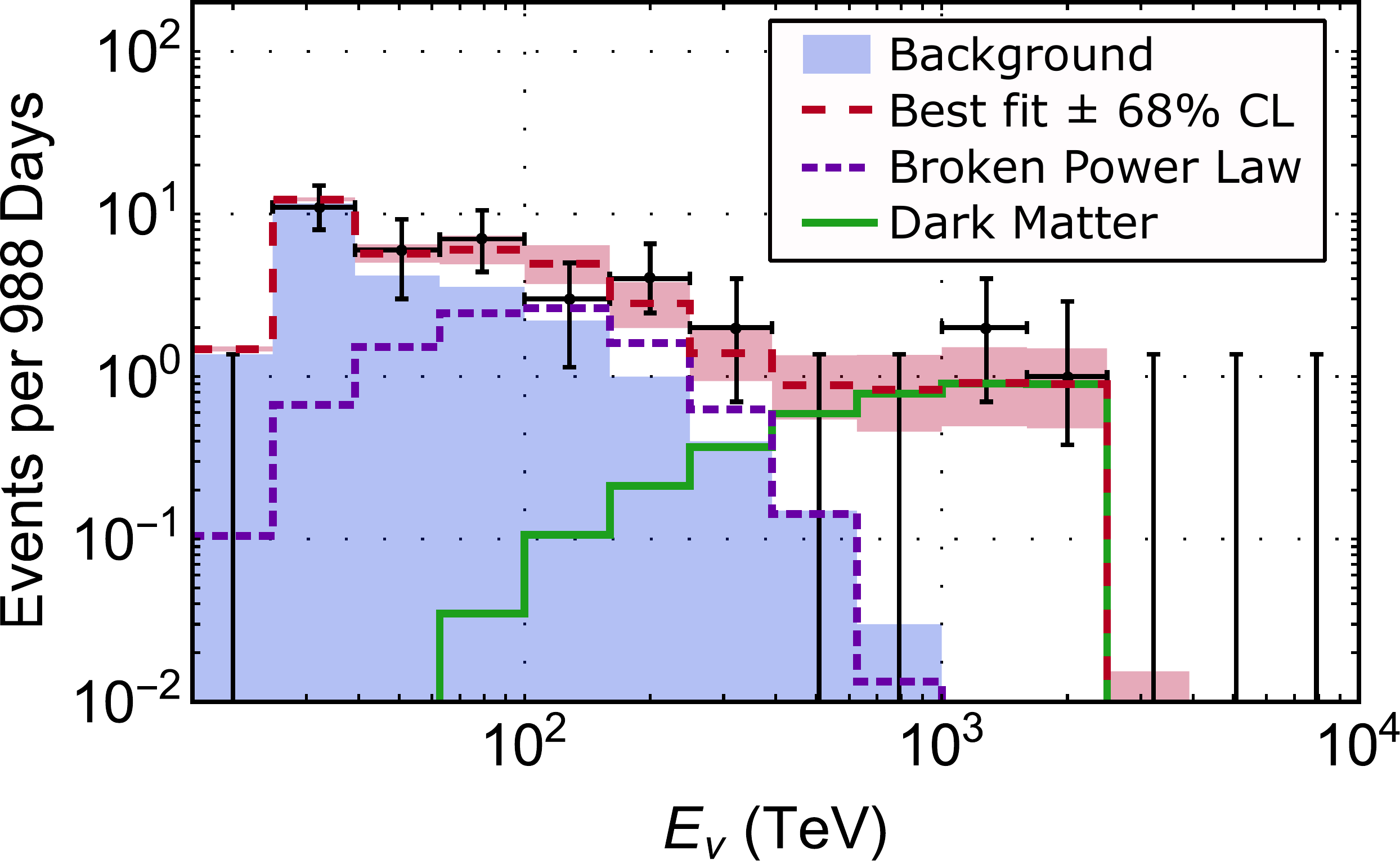}
\includegraphics[width=0.42\textwidth]{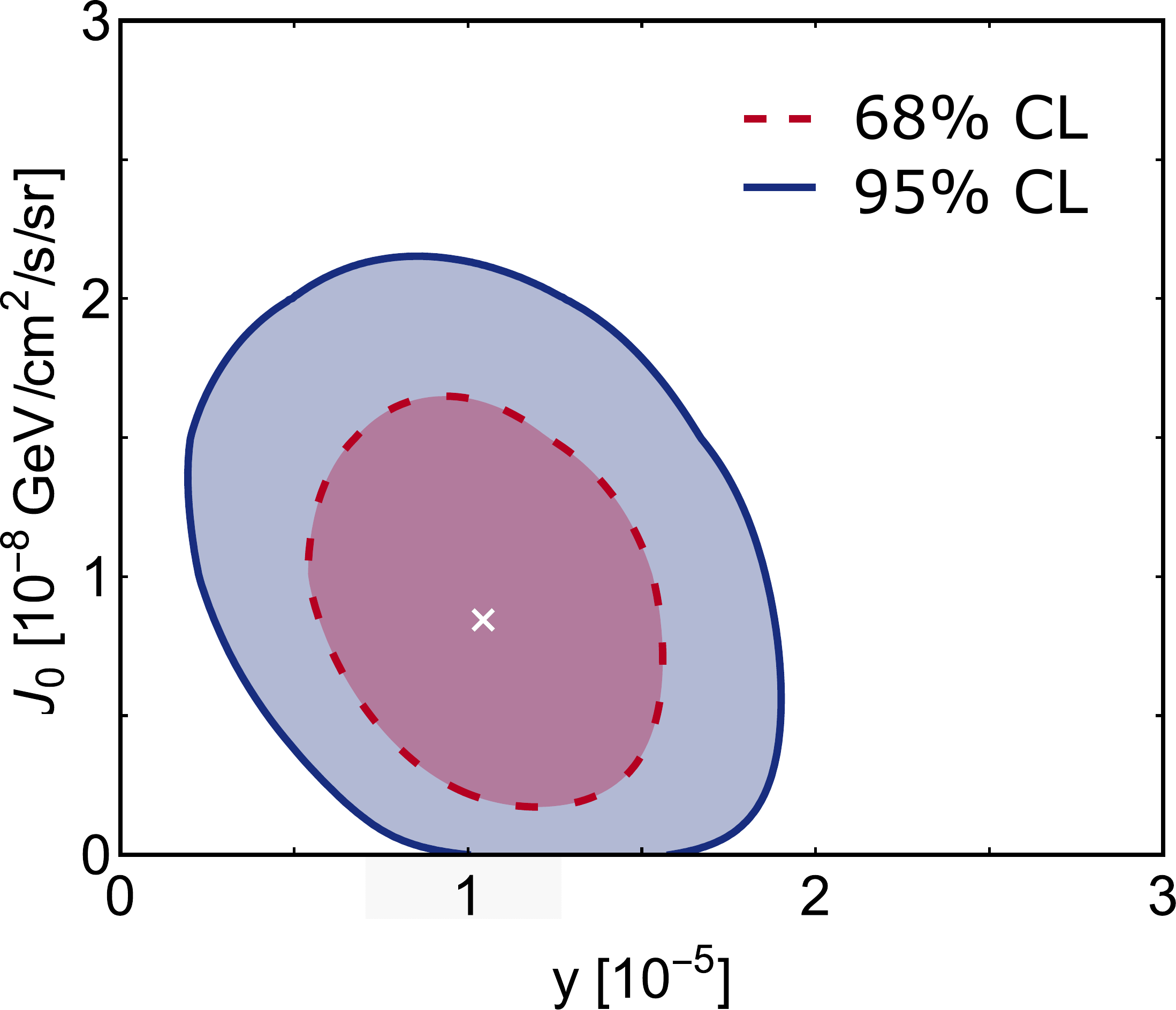}
\hskip5.mm
\includegraphics[width=0.42\textwidth]{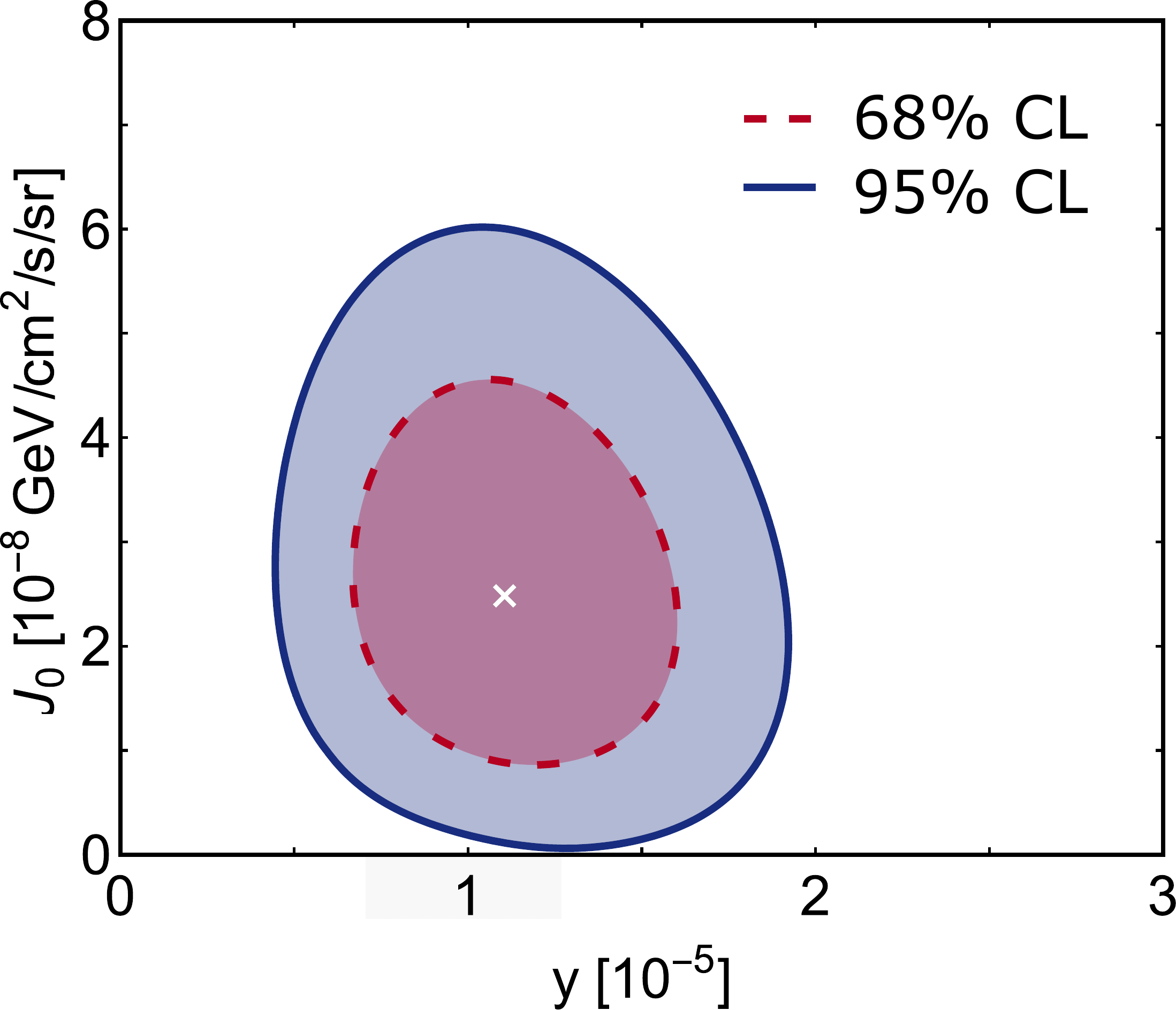}
\\(A)\hskip69.mm (B)\\
\caption{Results of the analysis for model 1). First row shows the neutrino events as a function of the neutrino energy $E_\nu$ for the DM + UPL (column A)  and  DM+BPL (column B) models. The red (long-dashed) line is the best fit (background + astrophysical + DM components), and its band represents the 68\% C.L. resulting from the fit. The purple (dashed) and green (solid) lines are the astrophysical and DM contributions, respectively. The black points are the IC data, and the blue region shows the upper limit for the sum of all backgrounds (see Ref.~\cite{Aartsen:2014gkd}). In the second row we report the 68\% C.L. (dashed) and 95\% C.L. (solid) contours for the two parameters $y$ and $J_0$ corresponding to DM + UPL (column A)  and  DM+BPL (column B). The crosses are the best-fit points.}
\label{fig:U1}
\end{center}
\end{figure}

In Table~\ref{tab2}, we give the marginalized 95\% C.L. for parameters $y$ and $J_0$, for models 1) and 2), and UPL and BPL parameterizations, respectively. From the values of reduced $\chi^2$ shown in the last column, we see  that the experimental data slightly prefer the BPL scheme with respect to the UPL one. In each case, the analysis shows that a non-vanishing DM contribution at 2-$\sigma$ level (95$\%$ C.L. for $y$ not compatible with zero) is required. This is mainly due to the presence of the sharp cutoff in the data at high energy. By comparing the results obtained for $U_f(1)$ and $A_4$, one cannot appreciate a significative difference. Indeed, the two models essentially provide similar features at the level of produced neutrino flux. For the cases reported in Table~\ref{tab2} we have checked that the total energy injected by DM decay in the e.m. sector is smaller than the bound provided by Fermi-LAT \cite{Ackermann:2014usa}.
\begin{figure}[t!]
\begin{center}
\includegraphics[width=0.46\textwidth]{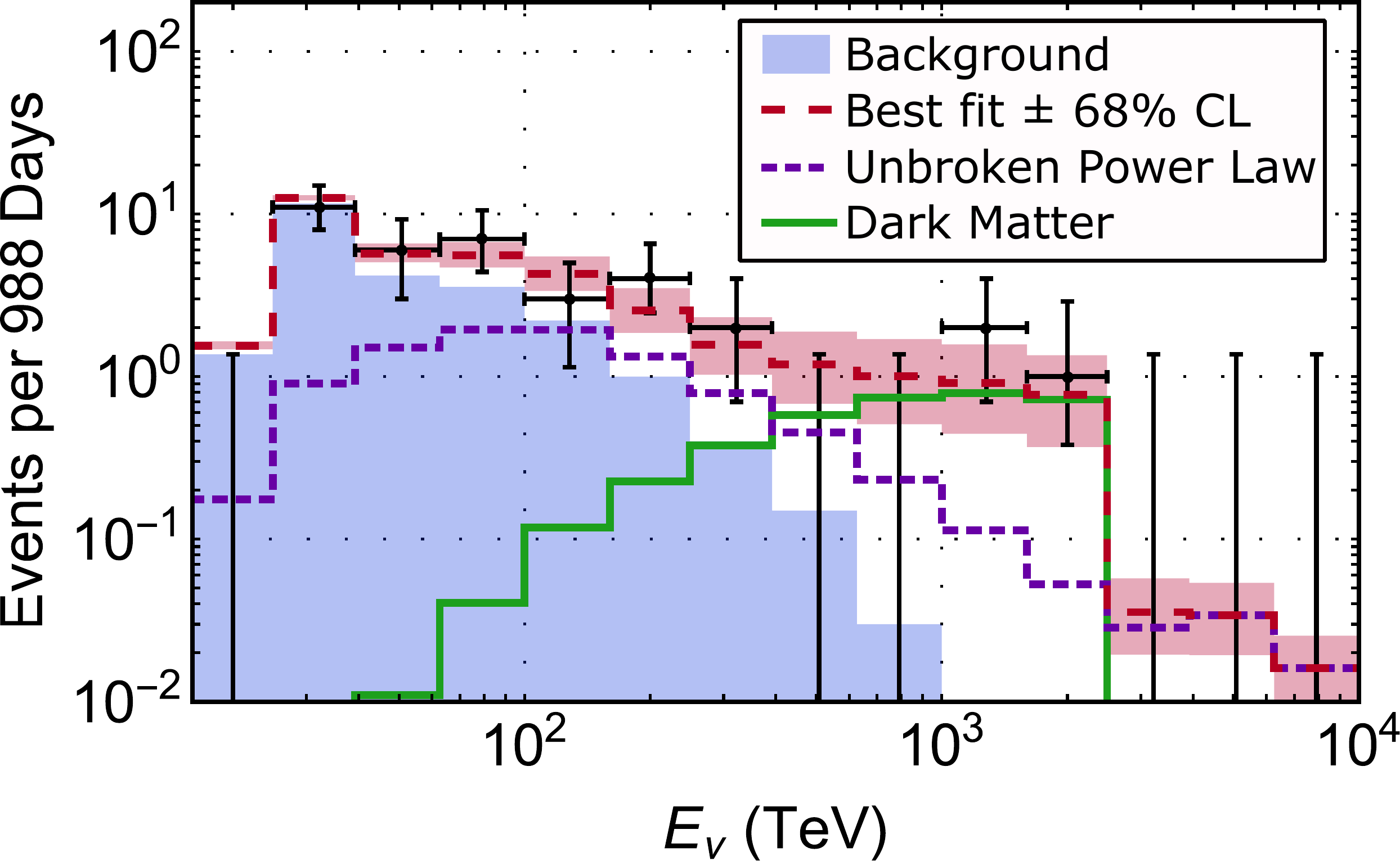}
\hskip5.mm
\includegraphics[width=0.46\textwidth]{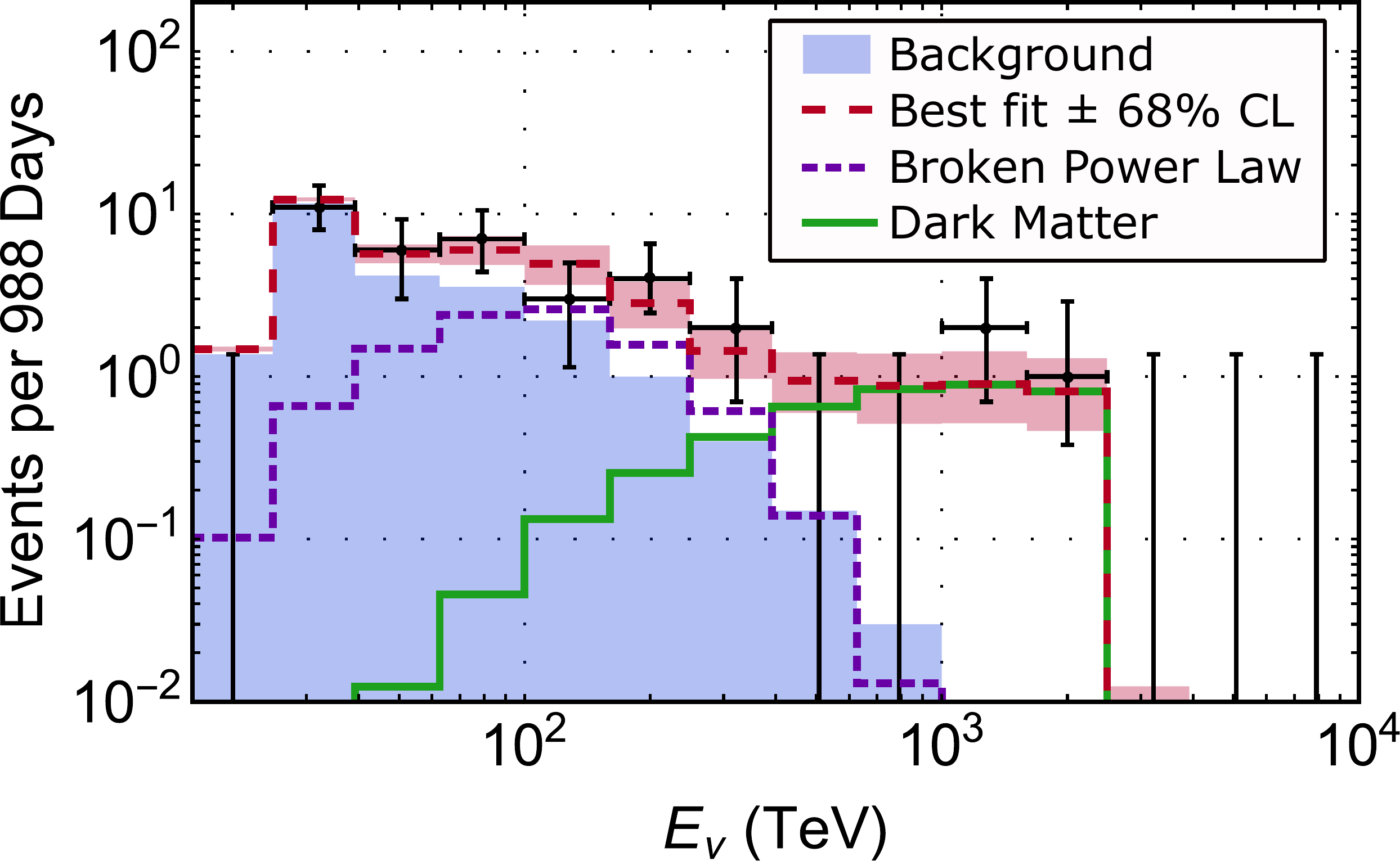}
\includegraphics[width=0.42\textwidth]{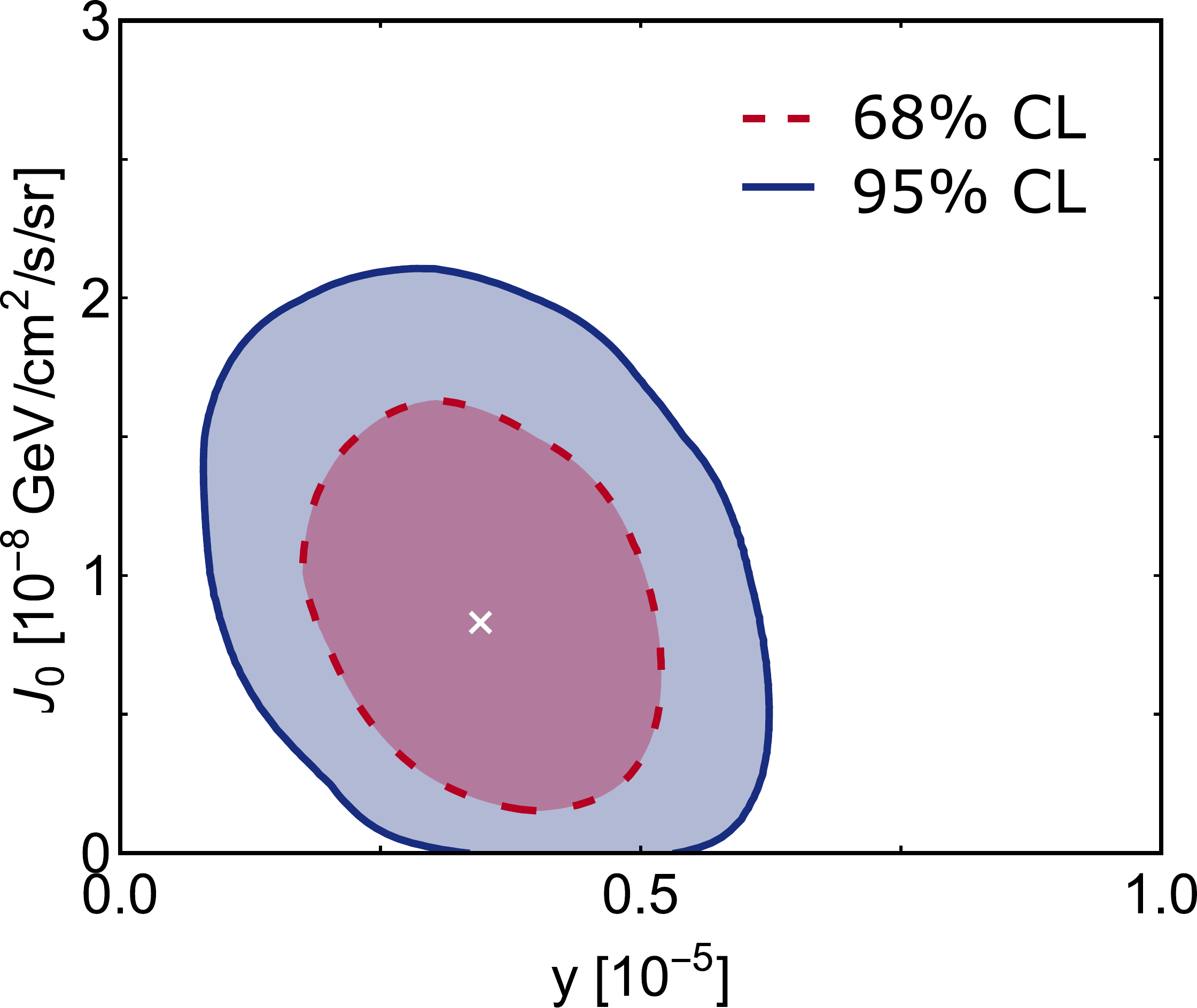}
\hskip5.mm
\includegraphics[width=0.42\textwidth]{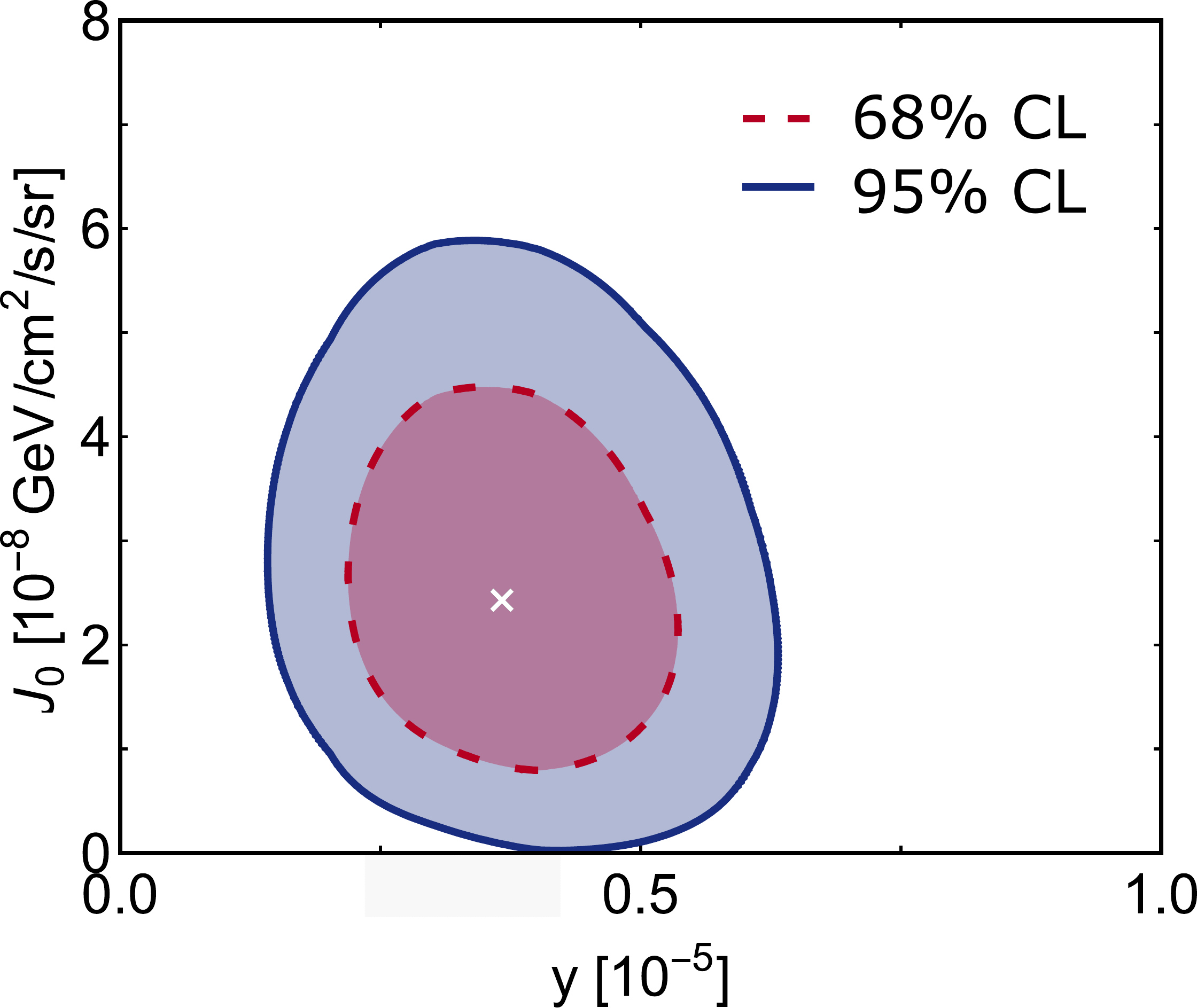}
\\(A)\hskip69.mm (B)\\
\caption{Same as Fig.~\ref{fig:U1} for model 2).}
\label{fig:A4}
\end{center}
\end{figure}
The neutrino events as function of $E_\nu$ for model 1) are shown in Fig.~\ref{fig:U1} (first row) and compared with IC data for the two models, DM + UPL (column A)  and  DM+BPL (column B). In the second row we report the 68\% C.L. (dashed) and 95\% C.L. (solid) contours for the two parameters, $y$ (defined in Eq.~\eqref{eq_def_y_1}) and $J_0$ (defined in Eq.s~\eqref{eq_UPL_def} \eqref{eq_BPL_def}), in the cases DM + UPL (column A)  and  DM+BPL (column B). The crosses stand for the best-fit points. The same quantities, but corresponding to the $A_4$ model, are reported in Fig.~\ref{fig:A4}. The plots in Figs.~\ref{fig:U1} and \ref{fig:A4} show that there is no significative difference between the two models: both two models predict the observation of neutrinos in the energy range [0.3 PeV, 1.0 PeV] and a sharp cutoff at the energy of few PeV. Notice that since in both $U_f(1)$ and $A_4$ symmetry cases the galactic and extragalactic components of the DM neutrino flux are of the same order of magnitude, we expect at the PeV energies an almost isotropic neutrino flux with a significant level of anisotropy near the galactic center. This is in a good agreement with the IC observations.
\begin{figure}[t!]
\begin{center}
\includegraphics[width=0.36\textwidth]{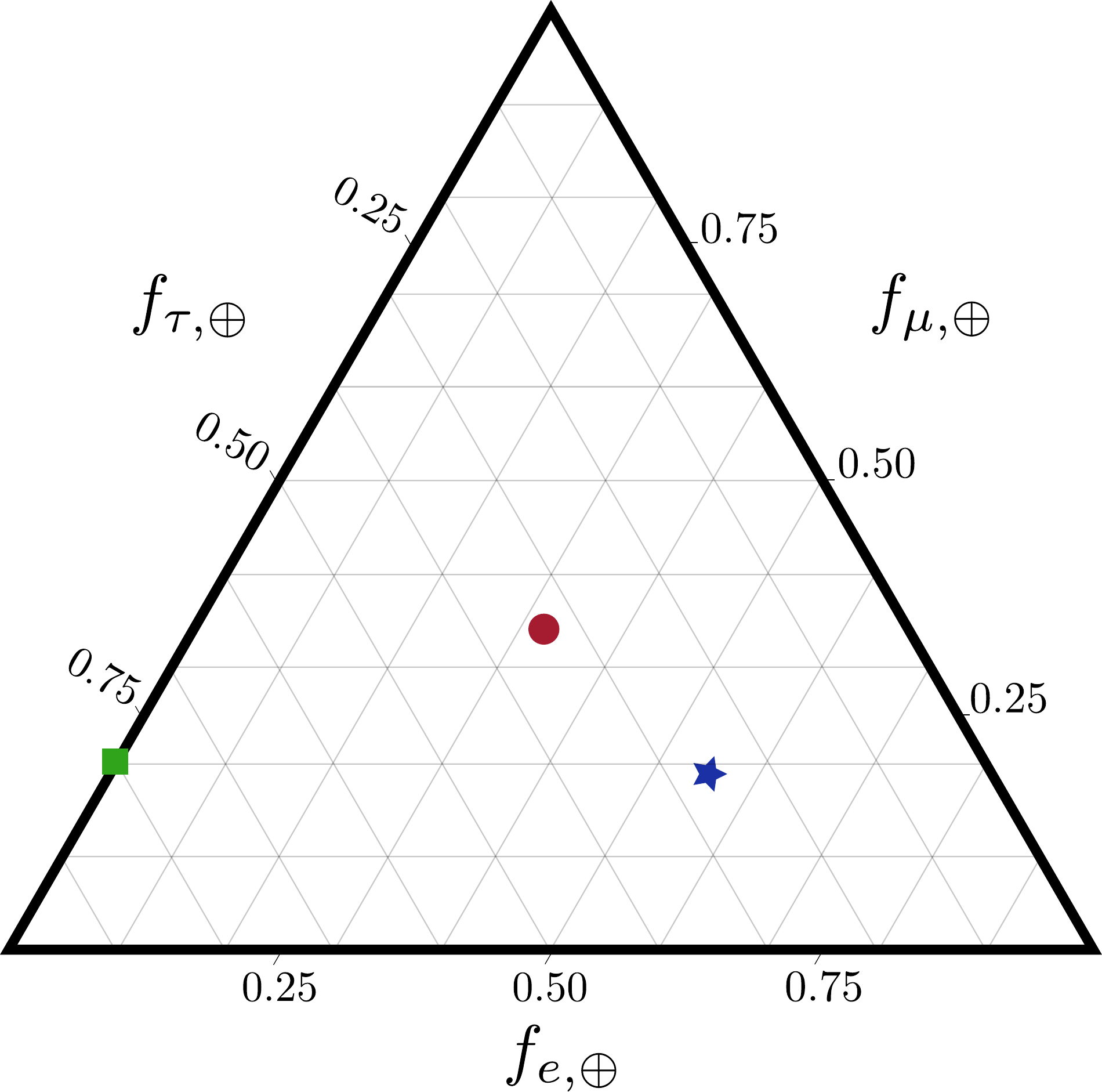}
\hskip5.mm
\includegraphics[width=0.52\textwidth]{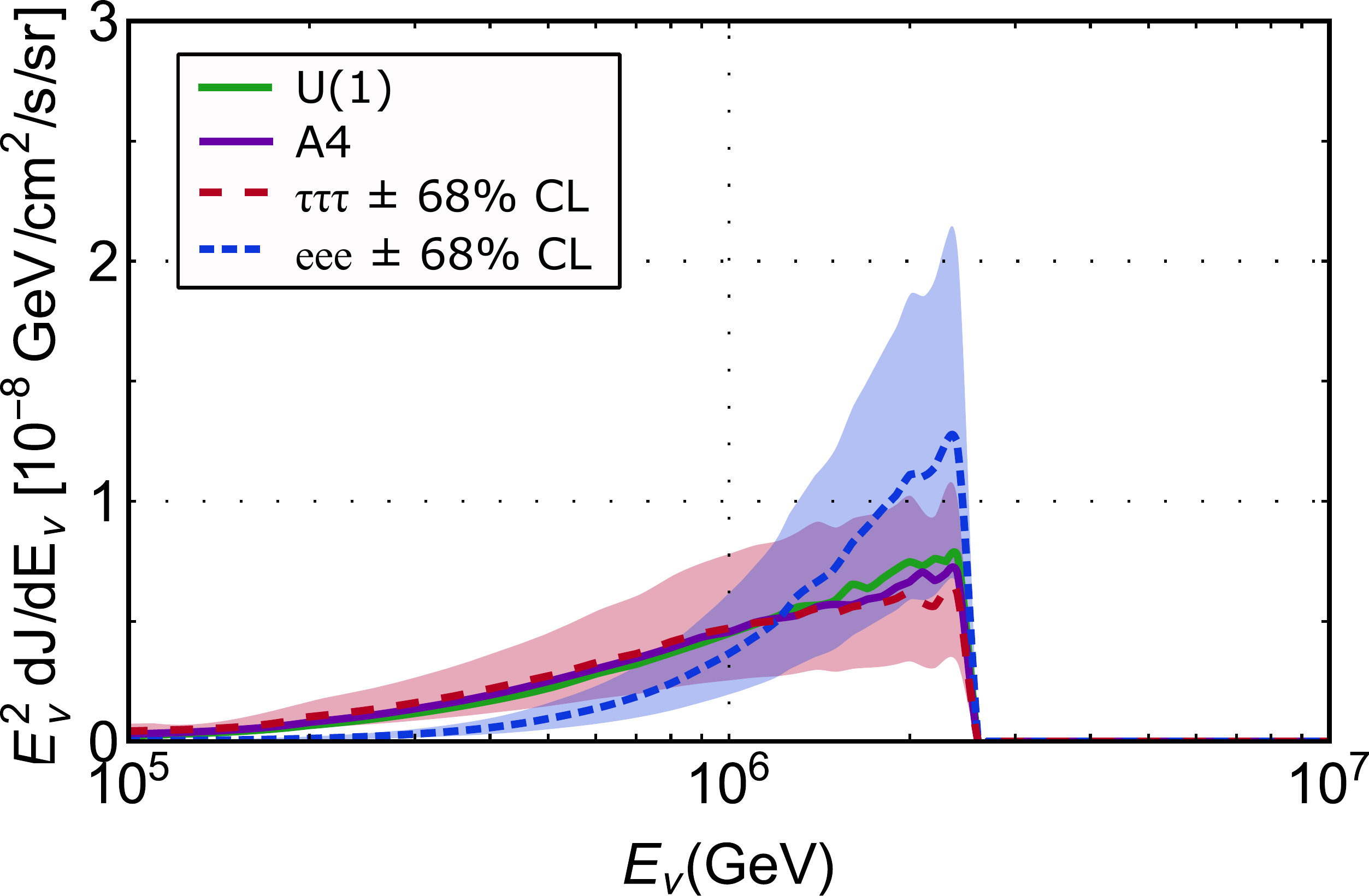}
\caption{Left panel shows the flavour compositions at Earth of the DM neutrino flux for model 1), 2) and the two fully diagonal cases $(e,e,e)$ and $(\tau,\tau,\tau)$, as well (see text). The green square represents the IC flavour analysis of Ref. \cite{Aartsen:2015ivb}, but referred to the integrated neutrino flux above 35 TeV. The prediction for models 1), and 2) and the $(\tau,\tau,\tau)$ case are represented by the red disk, whereas the blue star stands for the $(e,e,e)$ case. Right panel presents the DM neutrino flux for model 1) (green, solid), model 2) (purple, solid),  $(e,e,e)$ (blue, dashed), and $(\tau,\tau,\tau)$ (red, long-dashed). The two 68~$\%$ C.L. bands refer to the two cases $(e,e,e)$ and $(\tau,\tau,\tau)$.}
\label{fig:models}
\end{center}
\end{figure}

In order to understand the reason for the strong similarity between the predictions of model 1) and 2), we have also considered two situations where the operator of Eq.~\eqref{eq:DMop} is characterized by a single term only, $\{\alpha,\beta,\gamma\}\equiv \{ e,e,e \}$ and  $\{\alpha,\beta,\gamma\}\equiv \{\tau,\tau,\tau \}$, though they cannot be realized by using the flavour symmetries considered here, as shown in the Appendix. From a {\it phenomenological} point of view it is interesting to consider such cases because they correspond to the extreme situations with respect to the number of $\tau$ leptons produced and thus, of secondary neutrinos from their decays. In the left panel of Fig.~\ref{fig:models} we report the flavour compositions at Earth of the DM neutrino flux for model 1) and 2) and the two fully diagonal cases $(e,e,e)$ and $(\tau,\tau,\tau)$, as well. The black dot represents the IC flavour analysis of Ref.~\cite{Aartsen:2015ivb}. Models 1) and 2) and the $(\tau,\tau,\tau)$ diagonal case provide the same flavour composition at Earth, namely $\left(f_e : f_\mu : f_\tau \right)_\oplus \approx \left( 1/3 : 1/3 : 1/3 \right)$, which are the flavour ratios provided by the standard astrophysical sources. A different situation is given by the diagonal case $(e,e,e)$ where we have only electron (anti)neutrinos as DM decay products, that leads at Earth $\left(f_e : f_\mu : f_\tau \right)_\oplus \approx \left( 0.55 : 0.19 : 0.26 \right)$.  It is worth reminding that a study of shower/track composition of the PeV events can shed light on the flavour ratios and hence on the flavour structure of the coupling. Such analysis cannot be performed at the moment due to the low statistics available. In the right panel of Fig.~\ref{fig:models} one can see that changing the flavour structure of the DM-SM coupling does not appreciably affect the DM neutrino flux. 

\section{\bf Conclusions}
\label{sec:concl}

During the last three years of observation the IceCube Collaboration has observed neutrino events in the TeV--PeV range. Even though lower energy events can be well explained in terms of atmospheric neutrinos, the most energetic processes detected, in the range between 100~TeV and 2~PeV, seem to originate from some extraterrestrial source.  In the present analysis, by following the results on the background presented in Ref.~\cite{Aartsen:2014gkd}, we assumed an extraterrestrial neutrino flux dominating for energies larger than 60 TeV. In particular,  in order to recover the drop of the flux observed around few PeV, this high energy flux is obtained as the sum of two different components: a bottom-up neutrino contribution coming from known astrophysical sources (like for example extragalactic SNR) in the energy region [60~TeV, 300~TeV], and a top-down additional component, dominating at higher energies, originated from the decay of DM particles with mass of few PeV. The astrophysical flux was parametrized  using either an unbroken power law or a broken power law with an exponential cutoff.  We have considered both the options UPL and BPL  in order to cover the wide range of possible astrophysical sources. The top-down term in the neutrino flux is produced by the decay of a PeV DM particle, $\chi$, which we assume for simplicity to be the dominant  cold DM term. The decay of $\chi$ has been calculated by means of a MonteCarlo procedure.

A similar approach was already investigated in literature \cite{Esmaili:2014rma}, but assuming the presence of a trilinear coupling for $\chi$ like the one reported in Eq.~\eqref{eqLnC}. In this case, the $\chi$-decay induces a shower whose hadronic content yields secondary neutrinos that provide an almost flat behavior of neutrino spectrum at low energy. The main problem of such an approach is that it could be in contrast with the standard astrophysical sources since the trilinear interaction term alone totally explains the IC extraterrestrial neutrino flux. Moreover, in this scenario an {\it unnatural} tiny coupling for the interaction term of Eq.~\eqref{eqLnC}, $\mathcal{O}(10^{-30})$, is required. In the present analysis we use the same approach of \cite{Esmaili:2014rma}, but assume the presence of special flavour symmetries (Abelian and non-Abelian), like $U_f(1)$ and $A_4$ for example. Such symmetries forbid a trilinear interaction term of the type in Eq.~\eqref{eqLnC}, and allow quadrilinear leptophilic interaction terms only, like the one in Eq.~\eqref{eq:DMop}. At the same time they reproduce the values of neutrino masses and mixing. The leptophilic characteristic is crucial in forbidding the production of a huge hadronic component in the decay shower that would flatten again the neutrino spectrum. Such a model allows one to improve the problem of naturalness of the coupling, since the non-renormalizable terms is weighted by the square of the  new physics mass scale, that can be as large as the Planck mass. In this scenario the coupling required to recover the IC data results to be 25 orders of magnitude larger than the one for the trilinear term.

The comparison of the prediction for the neutrino events in UPL + DM and BPL + DM with the experimental observations allows the determination via a multi-Poisson likelihood approach of the free parameters of the model. In Table~\ref{tab2} we show the resulting parameters for the $U_f(1)$ and $A_4$ symmetry models. Both schemes provide a fair description of the experimental situation within the statistical uncertainty, with the BPL parameterizations providing a slightly smaller value of the reduced $\chi^2$. The contour plots seem to suggest at 2-$\sigma$ the presence of a hard component (top-down term) in the spectrum ($y$ not compatible with zero), whereas the astrophysical contribution could be even vanishing ($J_0$ compatible with zero). This is almost expected because the astrophysical contribution partially overlap with the atmospheric background (affected by a large uncertainty) up to few hundreds TeV. Nevertheless, we cannot conclude at this level of statistics that the presence of DM is really demanding. A confirmation by future data of the cutoff above few PeV is therefore particularly important. Moreover, since the galactic and extragalactic components of the DM neutrino flux are found to be of the same order of magnitude, another evidence in favor of a DM component would be the presence of a significant level of anisotropy near the Galactic Center. 

We also analyzed the flavour ratios of neutrinos at Earth predicted by the two models, and found that for both models, $\left(f_e : f_\mu : f_\tau \right)_\oplus \approx \left( 1/3 : 1/3 : 1/3 \right)$. Only a coupling of $\chi$ with first lepton generation only would produce clear features in the flavour ratios, which in principle could be measurable if a large statistics of events were available.

Finally, we would like to highlight that our analysis regards only the three years IC data. The recent new IC data provide a track event with a deposited energy of about 2.6 PeV, that could be explained by increasing the mass of DM particle in our model. However, since such an event is not fully contained, the energy of the primary neutrino is unknown and could also be very high ($\mathcal{O}(10)$ PeV). Note that if the energy of the primary neutrino was of the order of few PeV this would not change our conclusions, differently a neutrino energy of the order of 10 PeV would be in tension with our analysis.

\acknowledgments

We acknowledge support of the MIUR grant for the Research Projects of National Interest PRIN 2012 No. 2012CP- PYP7 {\it Astroparticle Physics} and of INFN  I.S. {\it TASP} 2014. The authors would like to thank Dr. S. Chakraborty for useful discussions. S.M.B. acknowledges support of the Spanish MICINN's Consolider-Ingenio 2010 Programme under grant MultiDark CSD2009-00064.

\section*{A Flavour symmetry schemes}

As discussed in \secs{sec:th}, we look for symmetry motivations able to single out the operator  of \Eq{DMop} among those listed in Table \ref{op}. The flavour symmetry schemes that can be used may be Abelian or non-Abelian. We will discuss in the following some possible benchmark models as relevant examples.
\begin{table}[t!]
\centering
\begin{tabular}{|c|cccc|c|}
\hline
 & $L_e, \ell_e$ & $L_\mu, \ell_\mu$ & $L_\tau, \ell_\tau$ & $\phi$ & $\chi$ \\ 
\hline
$q_\psi$ & 2 & 1 & 4 & 0 & 3 \\ 
\hline
\end{tabular}
\vskip5.mm
\begin{tabular}{|c|ccc|c|}
\hline
& $L$ & $\ell$ & $\phi$ & $\chi$ \\ 
\hline
$A_4$ & ${\bf 3}$ & ${\bf 3}$ &  ${\bf 1}$ & ${\bf 1}$ \\ 
\hline
\end{tabular}
\caption{Charges, $q_\psi$, for the $U_f(1)$ symmetry (upper Table). In the lower Table we report the Irreducible Representations allocating SM fields and the dark matter $\chi$ in the non-Abelian $A_4$ symmetric model (quarks are all singlets under $A_4$).}
\label{tab:U1}
\end{table}

\section*{B Abelian case}

Let us denote with $q_\psi$  the $U_f(1)$ flavour charge of a generic field $\psi$. It can be shown that  a flavour Abelian symmetry cannot single out a flavour-diagonal operator in Eq.~\eqref{eq:DMop}. In fact, in this case both the operators $\bar{L}_\alpha {\ell}_\alpha  \bar{L}_\alpha   \chi$ and $\bar{L}_\alpha \phi \ell_\alpha$ should be invariant  under such a symmetry, but this would imply the invariance 
of the term $L_\alpha \tilde{\phi}  \chi$ as well. Indeed,  in terms of abelian charges we would have $-2q_{L_\alpha}+q_{\ell_\alpha}+q_\chi=0$ and $-q_{L_\alpha}+q_\phi+q_{\ell_\alpha}=0$, which implies $-q_{L_\alpha}-q_\phi+q_\chi=0$.\footnote{Note however that this conclusion could be bypassed by invoking supersymmetry.} However, if we mix different lepton flavours then we have more freedom to consistently define the charges of a $U_f(1)$ flavour symmetry. This can be done in such a way that only a leptophilic dimension 6 operator of Table \ref{op} results to be invariant.  A possible realization of such a scheme is shown in Table \ref{tab:U1}. In this case the only invariant DM decay operators are 
\begin{equation}
\label{eq:DMopU1}
\mathcal{O}_6=\frac{1}{M_{\rm Pl}^2} \left( y_{\mu e \tau} \,\overline{L_\mu}{\ell}_e \overline{L_\tau}  \chi + y_{\tau e \mu} \overline{L_\tau}{\ell}_e \overline{L_\mu}  \chi + y_{e \mu e} \,\overline{L_e}{\ell}_\mu \overline{L_e}  \chi \right)+ \mathrm{h.c.} \ .
\end{equation}
The charge assignment in Table~\ref{tab:U1} is by no means unique; different choices would select different operators. Moreover with the charge assignment given in Table~\ref{tab:U1}, the dimension five Weinberg operator $\overline{L_\alpha}L_\beta^c \tilde{\phi}\tilde{\phi}$ has a structure very similar to the so-called $B_4$ two-zeros texture given in \cite{Frampton:2002yf} that can fit the lepton mixing parameters, see for instance \cite{Fritzsch:2011qv,Ludl:2011vv}.

In principle the form of the operator is dictated by experiment. A hypothetical study of the flavour composition of PeV neutrinos in IC data would give useful hints in the definition of the possible flavour charges.

\section*{C Non-Abelian case}

Another possible realization of our scheme invokes non-Abelian discrete symmetries. A simple model would employ for instance $\Delta(27)$ by assigning $L$, $\phi$ and $\ell$ to $\mathbf{3}$ Irreducible Representation while $\chi$ remains blind to the symmetry. However, for the sake of definiteness we will consider in the rest of this article a framework based on $A_4$ flavour symmetry with $L$ and $\ell$ transforming as $\mathbf{3}$ while all the remaining fields are singlets (for more details about such a model we refer to \cite{Haba:2010ag}). Our conclusions are not affected by this choice. The Lagrangian of the model contains the following relevant terms:
\begin{equation}
\label{eq:lag}
\mathcal{L} \supset -\frac{1}{2}(M_\chi\, \overline{\chi^c} \chi+\,\mathrm{h.c.}) +\mathcal{O}_{6}  \ ,
\end{equation}
where $\mathcal{L}_{dim=6}$ is the lowest order non-renormalizable operator allowed by the symmetries of the model. It represents a Fermi-like decay interactions give by
\begin{equation}\label{eq:lag6}
\mathcal{O}_{6}= \frac{y}{M_{\rm Pl}^2} \,(\bar L \ell)_{3} \bar L \,\chi + \frac{y^\prime}{M_{\rm Pl}^2}\, (\bar{L} \ell)_{3'} \bar L \,\chi \ ,
\end{equation}
where the notation $(..)_{3(3')}$ reflects the fact that in $A_4$ there are two possible contractions of two triplets into one triplet representation. We see that Eq.~\eqref{eq:lag6} not only reproduces Eq.~\eqref{eq:DMop} but it also significantly simplifies its structure. Note that in this case we have two independent  couplings, namely $y$ and $y^\prime$.

\end{document}